\documentclass[useAMS,usenatbib]{mn2e}
\usepackage{graphicx}
\usepackage{color}
\usepackage{amssymb,amsmath}

\title[The physical properties of {\it Fermi} BL Lac objects jets]{The physical properties of {\it Fermi} BL Lac objects jets}
\author[Yan et al.]{Dahai Yan, Houdun Zeng, and Li Zhang\thanks{E-mail: lizhang@ynu.edu.cn}\\
Department of Physics, Yunnan University, Kunming, China}

\begin{document}

\date{Accepted 2014 January 18. Received 2013 October 27; in original form 2013 January 17.}

\pagerange{\pageref{firstpage}--\pageref{lastpage}} \pubyear{2013}

\maketitle

\label{firstpage}

\begin{abstract}
We investigate the physical properties of BL Lac objects (BL Lacs)
jets by modeling the quasi-simultaneous spectral energy
distributions (SEDs) of 22 {\it Fermi} BL Lac objects in the frame of
a simple one-zone synchrotron self-Compton (SSC) model. We obtained the best-fit
model parameters and their uncertainties for each BL Lac object through the
$\chi^2$-minimization procedure and discussed their implications on the physical
processes. The modeling results show that the one-zone SSC model can successfully fit the SEDs
of high-synchrotron-peaked BL Lacs (HBLs) and
intermediate-synchrotron-peaked BL Lacs (IBLs), but fails to explain the SEDs of low-synchrotron-peaked BL Lacs (LBLs).
The statistical analysis results for model parameters are summarized as follows.
(1) No correlation is found between magnetic field ($B$) and
the broken energy of relativistic electrons distribution
($\gamma^{\prime}_{\rm b}$) for HBLs and IBLs, but there are inverse correlations between $\gamma^{\prime}_{\rm
b}$ and the radius of emitting blob ($R^{\prime}_{\rm b}$) as well as the electrons number $K^{\prime}_{\rm e}$
for HBLs and IBLs. It's therefore concluded that the variation of
$\gamma^{\prime}_{\rm b}$ is mainly caused by that of
$R^{\prime}_{\rm b}$ rather than $B$ for HBLs and IBLs. (2) The
Poynting flux in jets can not account for the observed radiations
since the power in Poynting flux is smaller than the radiative
power, and the cold protons could be the
primary energy carrier in the jets.
\end{abstract}


\section{Introduction}

The observed emissions from blazars are dominated by the
non-thermal emissions from relativistic jets. Their spectral energy
distributions (SEDs) are characterized by two distinct bumps: the
first bump located at low-energy band is dominated by the
synchrotron emission of relativistic electrons, and the second
bump located at high-energy band could be produced by inverse
Compton (IC) scattering (e.g., B\"{o}ttcher 2007). The seed
photons for IC process could be from the local synchrotron
radiation of the same relativistic electrons (i.e. synchrotron
self-Compton (SSC); e.g., Rees 1967; Maraschi et al. 1992;
Tavecchio et al. 1998), or from the external photon fields (EC;
e.g., Ghisellini \& Tavacchio 2009; Dermer et al. 2009), such as
those from accretion disk (e.g., Dermer \& Schlickeiser 1993) and
broad-line region (e.g., Sikora et al. 1994). The hadronic model
is an alternative explanation for the high energy emissions from
blazars (e.g., Mannheim 1993; M\"{u}cke et al. 2003; Dimitrakoudis
et al. 2012; Dermer et al. 2012).

Blazars are traditionally divided into BL Lac objects (BL Lacs) and flat
spectrum radio quasars (FSRQs) based on the emission line
equivalent width (EW) being smaller or larger than 5\AA. In
general, the SEDs of BL Lacs having weak lines or absent lines can
be reproduced well by the homogenous one-zone SSC model
\citep[e.g.,][]{zhang12}, and the EC components are
needed to explain the observed
high energy radiation from FSRQs \citep[e.g.,][]{G10,G11,yan}.
However, it should be kept in mind that it seems that the one-zone
SSC model sometimes fails to explain the gamma-ray emissions of
several intermediate-synchrotron-peaked BL Lacs (IBLs; e.g., Abdo et
al. 2011a) and low-synchrotron-peaked BL Lacs (LBLs; e.g., Abdo et
al. 2011b), and even high-synchrotron peaked BL Lacs (HBLs; e.g.,
Aliu et al. 2012a).

The modeling of SED with a given radiation mechanism allow us to
investigate the intrinsic physical properties of emitting region
\citep[e.g.,][]{G98,G08,zhang12,yan,yan2,man11,man} and the physical
conditions of jet \citep[e.g.,][]{Celotti08,G09jet,G10,G11}, like
the compositions, energy carrier and radiative efficiency of the
jet. With a large number of blazars, \citet{Celotti08} suggested
that the jet should comprise a dominant proton component and only
a small fraction of the jet power is radiated if there is one
proton per relativistic electron. \citet{G11} found that there is
a positive correlation between the jet power and the accretion
disk luminosity for {\it Fermi} broad-line blazars, and confirmed
that the jet should be protons dominated. In these previous
studies, \citet{G09jet,G10,G11} mainly concerned the relation
between the jet power and the accretion disk luminosity in {\it
Fermi} blazars, which is more significant for FSRQs.
\citet{Celotti08} estimated the powers of blazars jets based on
EGRET observations.

Since blazars show violent variability in multi-frequency bands,
especially at the X-ray and gamma-ray bands, so the SEDs obtained
simultaneously or quasi-simultaneously are crucial to reveal the
physical properties of their jets. Moreover, to well constrain the model
parameters and obtain the robust results for a given blazar, a
high-quality SED with a good coverage is needed. Fortunately,
\citet{abdosed} have assembled high-quality multi-wavelength data
of 48 blazars in the first three months of the LAT sample (LAT
Bright AGN Sample: LBAS) to build their quasi-simultaneous SEDs.
The data from {\it Swift} were collected in one day, or several
days; however, the {\it Fermi}-LAT data have been averaged over a
period of three months. Therefore, the multi-frequency data are
quasi-simultaneous, but not really simultaneous.

Because the SEDs of BL Lacs suffer less contamination of the
emission from the accretion disk and EC process which always can
be explained well by the one-zone SSC model, in order to reduce the
uncertainties on the radiative mechanisms, we only consider BL
Lacs here. In this work, we mainly take advantage of the
quasi-simultaneous SEDs of BL Lacs having certain redshift reported in \citet{abdosed} to
study the physical properties of their jets systematically within
the frame of the simple one-zone SSC model. Moreover, we use the
Levenberg-Marquardt (LM) method of $\chi^2$-minimization fitting
procedure instead of the ``eyeball" fitting to find the best-fit values of model parameters
and their uncertainties. The cosmological parameters ($H_0, \Omega_m,
\Omega_{\Lambda}$) = (70 km s$^{-1}$ Mpc$^{-1}$, 0.3, 0.7) are
used throughout this paper.

\section{Modeling fitting procedure}

The one-zone SSC assumes that non-thermal radiation is produced by
both the synchrotron radiation and SSC process in a spherical blob
filled with the uniform magnetic field ($B$), moving
relativistically at a small angle to our line of sight, and the
observed radiation is strongly boosted by a relativistic Doppler
factor $\delta_{\rm D}$. The radius of emitting blob is
$R^{\prime}_{\rm b}=\frac{t_{\rm v,min} \delta_{\rm D} c}{1+z}$,
where $t_{\rm v,min}$ is the minimum variability timescale. In
this study, we assume a broken power-law electron energy
distribution in the blob, and use the relativistic electron
distribution given by \citet{dermer09}, i.e.,
\begin{eqnarray}
N_{\rm e}^{\prime}(\gamma^{\prime})=K_{\rm
e}^{\prime}H(\gamma^{\prime};\gamma_{\rm min}^{\prime},\gamma_{\rm
max}^{\prime})\{{\gamma^{\prime
-p_1}\exp(-\gamma^{\prime}/\gamma_{\rm b}^{\prime})}
\nonumber \\
\times H[(p_{\rm 2}-p_{\rm 1})\gamma_{\rm
b}^{\prime}-\gamma^{\prime}]+[(p_{\rm 2}-p_{\rm 1})\gamma_{\rm
b}^{\prime}]^{p_{\rm 2}-p_{\rm 1}}\gamma^{\prime -p_{\rm 2}}
\nonumber \\
\times \exp(p_{\rm 1}-p_{\rm 2})H[\gamma^{\prime}-(p_{\rm
2}-p_{\rm 1})\gamma_{\rm b}^{\prime}]\}\;, \label{Eq1}
\end{eqnarray}
where $K_{\rm e}^{\prime}$ is the normalization factor,
$\gamma'$ is the Lorentz factor of a relativistic
electron with rest mass $m_e$, and $c$ is the speed of light.
$H(x;x_{1},x_{2})$ is the Heaviside function: $H(x;x_{1},x_{2})=1$
for $x_{1}\leq x\leq x_{2}$ and $H(x;x_{1},x_{2})=0$ everywhere
else; as well as $H(x)=0$ for $x<0$ and $H(x)=1$ for $x\geq0$. The
minimum and maximum energies of electrons in the blob are
$\gamma_{\rm min}^{\prime}$ and $\gamma_{\rm max}^{\prime}$,
respectively. This spectrum is smoothly connected with indices $
p_1$ and $p_2$ below and above the electron's break energy
$\gamma_{\rm b}^{\prime}$. Here, quantities in the observer's
frame are unprimed, and quantities in the comoving frame are
primed. Note that the magnetic field $B$ is defined in the
comoving frame, despite being unprimed.

For a given BL Lac with the electron energy distribution given by
Eq. (\ref{Eq1}), the local non-thermal spectra is
\begin{equation}
 f_{\epsilon}^{\rm tot}=f^{\rm syn}_{\epsilon}+f^{\rm SSC}_{\epsilon}\;.\label{Eq2}
\end{equation}
In the right hand of Eq. (\ref{Eq2}), the first term represents
the synchrotron spectrum, which can be given by Finke et al. (2008)
\begin{equation}
f^{\rm syn}_{\epsilon}=\frac{\sqrt{3}\delta^4_{\rm
D}\epsilon'e^3B}{4\pi h d^2_{\rm
L}}\int^\infty_0d\gamma'N'_e(\gamma')R(x)\;\;, \label{Eq3}
\end{equation}
where $e$ is the electron charge, $B$ is the magnetic field
strength, $h$ is the Planck constant, $d_{\rm L}$ is the distance
to the source with a redshift $z$,
$\epsilon^{\prime}=[h\nu(1+z)/m_{\rm e}c^2]/\delta_{\rm D}$ is
synchrotron photons energy in the co-moving frame. In equation
(\ref{Eq3}),
$R(x)=(x/2)\int^\pi_0d\theta\sin\theta\int^\infty_{x/\sin\theta}dt
K_{5/3}dt$, where $x=4\pi\epsilon'm^2_ec^3/3eBh\gamma^{'2}$,
$\theta$ is the angle between magnetic field and velocity of high
energy electrons, and $K_{5/3}(t)$ is the modified Bessel function
of order 5/3. Here we use an approximation for
$R(x)$ given by Finke et al. (2008).
The second term of the right hand of Eq.
(\ref{Eq2}) represents the SSC spectrum and is for isotropic and
homogeneous photon and electron distributions (e.g., Finke et al.
2008)
\begin{eqnarray}
f^{\rm SSC}_{\epsilon_{\rm s}}= \frac{9}{16} \frac{ (1+z)^2\sigma_{\rm T} \epsilon^{\prime 2}_s}
{\pi\delta^2_{\rm D} c^2 t_{\rm v,min}^2 } \int^\infty_0\ d\epsilon^{\prime}\
\frac{f_{\epsilon}^{\rm syn}}{\epsilon^{\prime 3}}\\\nonumber \times\int^{\gamma^{\prime}_{\rm max}}_{\gamma^{\prime}_{\rm min}}\ d\gamma^{\prime}\
\frac{N^{\prime}_e(\gamma^{\prime})}{\gamma^{\prime 2}} F_C(q^{\prime},\Gamma^{\prime}_{\rm e})\ ,\label{Eq4}
\end{eqnarray}
where $\sigma_{\rm T}$ is the Thomson cross section, $m_{\rm
e}c^2\epsilon^{\prime}_{s}=h\nu(1+z)/\delta_{\rm D}$ is the energy
of IC scattered photons in the co-moving frame, $F_{\rm
C}(q^{\prime},\Gamma_{\rm e}^{\prime})=2q^{\prime}{\rm
ln}q^{\prime}+(1+2q^{\prime})(1-q^{\prime})+\frac{q^{\prime
2}\Gamma_{\rm e}^{\prime 2}}{2(1+q^{\prime}\Gamma_{\rm
e}^{\prime})}(1-q^{\prime})$, $
q^{\prime}=\frac{\epsilon^{\prime}/\gamma^{\prime}}{\Gamma^{\prime}_{\rm
e}(1-\epsilon^{\prime}/\gamma^{\prime})}$, $\Gamma_{\rm
e}^{\prime}=4\epsilon^{\prime}\gamma^{\prime}$, and $
\frac{1}{4\gamma^{\prime 2}}\leq q^{\prime}\leq1$.

In this model, there are nine free parameters. Six of them specify
the electron energy distribution ($K^{\prime}_e$,
$\gamma^{\prime}_{\rm min}$, $\gamma^{\prime}_{b}$,
$\gamma^{\prime}_{\rm max}$, $p_1$, $p_2$), and the other three
ones describe the global properties of the emitting region ($B$,
$R^{\prime}_{\rm b}$, $\delta_D$). For a given BL Lac, we will
calculate its non-thermal flux using Eq. (\ref{Eq2}) and fit the
observed multi-wavelength data using the LM algorithm given by
\citet{press}. Because the LM method requires the initial input
values of the model parameters, at beginning we do a preliminary
modeling to the SED for each object to guess the starting values
for parameters. More details about the applications of the LM
method can be found in \citet{man11,man}. However, it should be
stressed that there may be caveats involved in applying the
$\chi^2_{\rm \nu}$ fitting to the non-linear model such as the SSC
model \citet{caveat}. As pointed out by \citet{caveat}, for the
non-linear model the number of degrees of freedom is hard to
determined, consequently the reduced $\chi^2$ is not a good method
for model assessment and model comparison no longer. Other methods
are needed to make judgement of goodness of the fit and make model
comparison (e.g., applying the Kolmogorov - Smirnov (K-S) test for
normality of the residuals of the SED fits in \citet{man11} and
investigating the convergence of the model parameter in
\citet{yan2}). Nevertheless, minimising $\chi^2$ is the correct
thing in order to fit the model to the observed data
\citep{caveat}. In this work, the LM method is just used to obtain
the best-fit model parameters and their uncertainties. The uncertainties returned by LM method is the approximated symmetrical (standard) errors (the square root of the covariance matrix diagonal elements) of the model parameters, which depend on a quadratic approximation to the $\chi^2$-surface around the minimum \citep{press,man11}.

The low energy cutoff in electrons distribution $\gamma^{\prime}_{\rm min}$
is always poorly constrained by the SED modeling. In this
study, in order to avoid overproducing the radio flux we set $\gamma^{\prime}_{\rm min}=500$ for Mrk 421 and BL Lac and $\gamma^{\prime}_{\rm min}=200$ for S5 0716+714, PKS 0851+202, GB6 J1058+5628 as well as ON 231 (W comae), and $\gamma^{\prime}_{\rm min}=100$ for the rest of objects.

\section{Applications}

\begin{figure*}
\includegraphics[width=15cm,height=14cm]{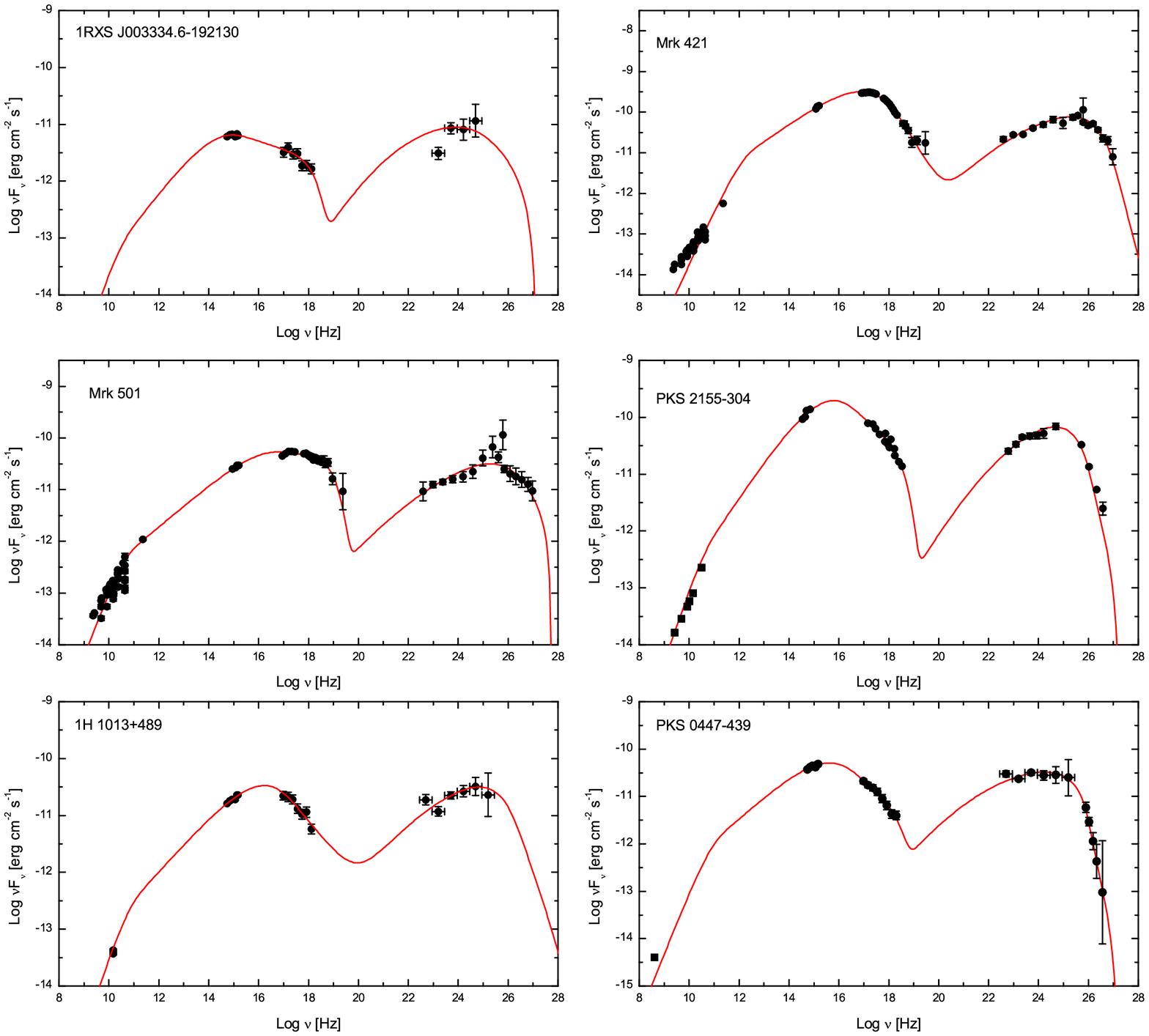}
\includegraphics[width=15cm,height=9.0cm]{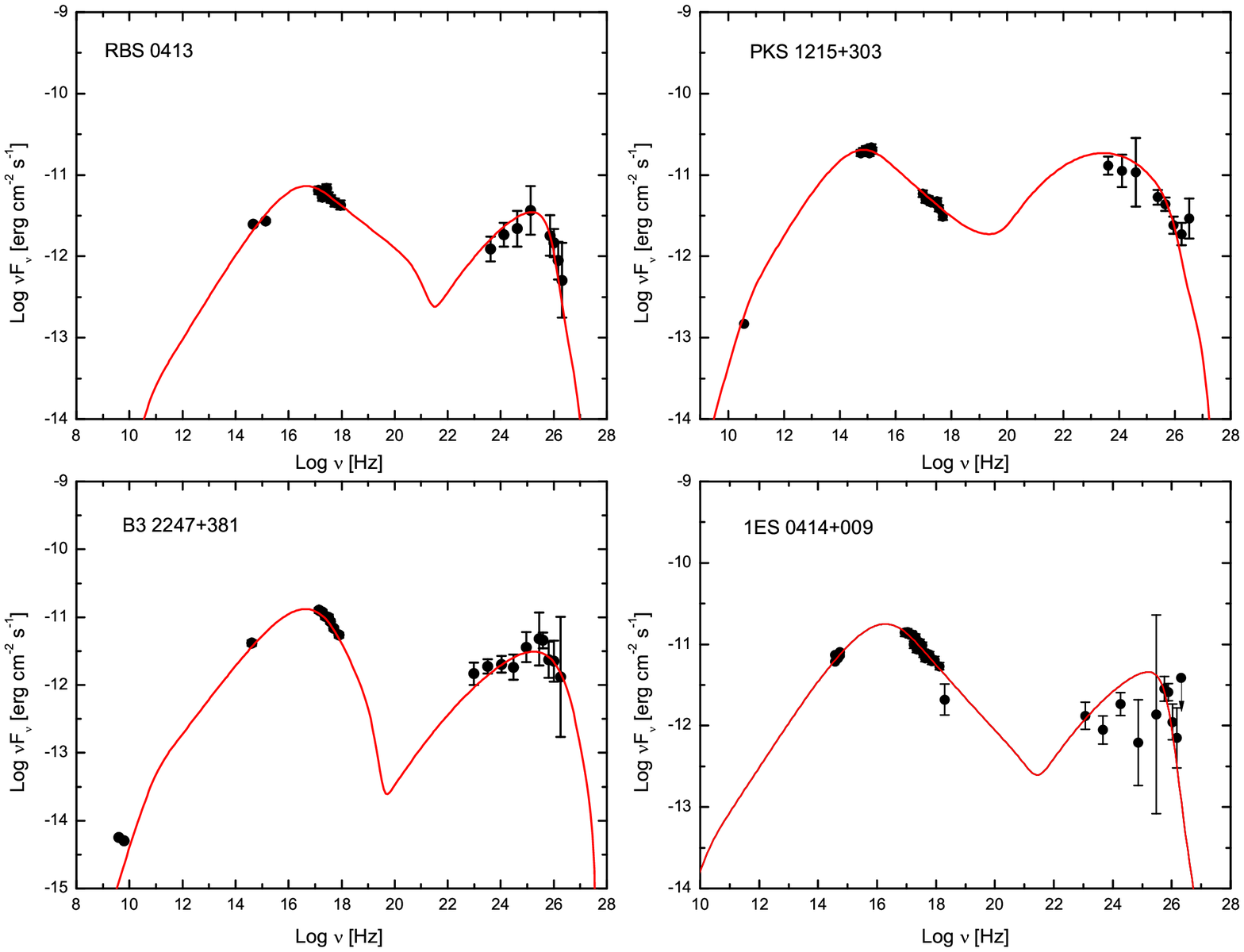}
\caption{Best-fit one-zone SSC modeling for the observed SEDs of HBLs.
}
\label{f1}
\end{figure*}

\begin{figure*}
\includegraphics[width=18cm,height=15cm]{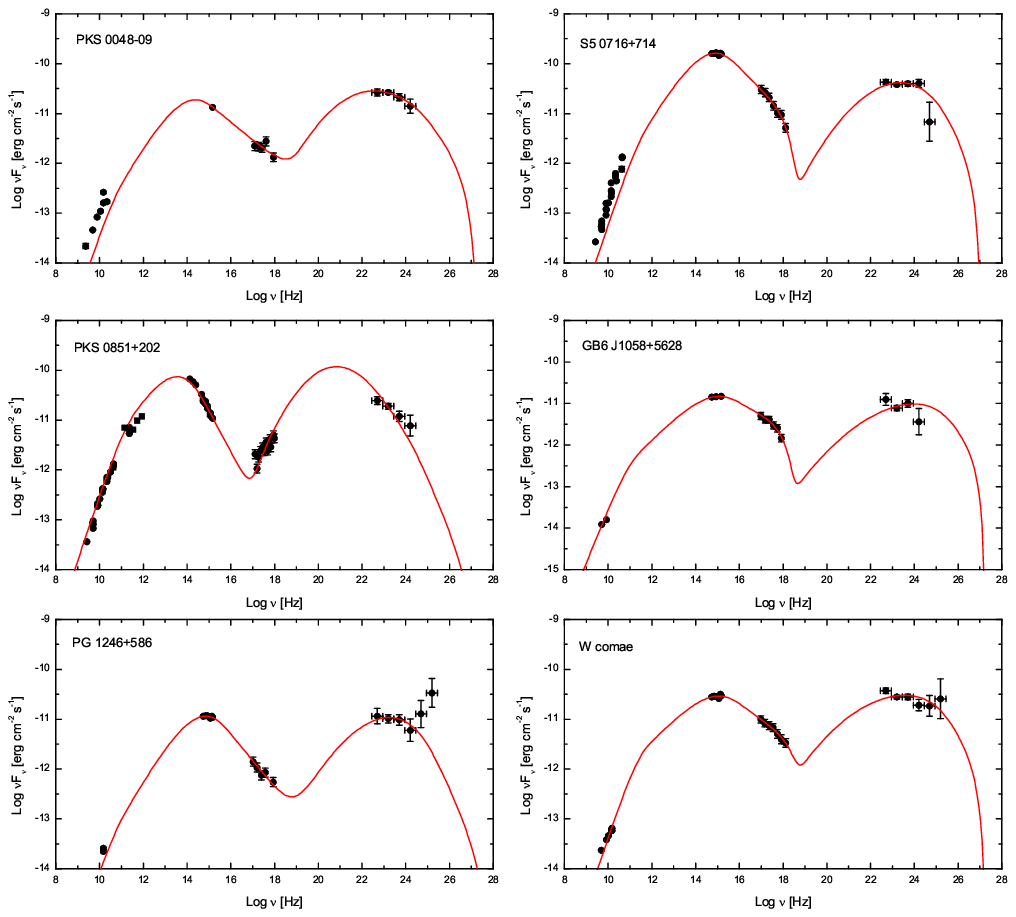}
\caption{Same as in Figure~\ref{f1}, but for IBLs.
}
\label{f2}
\end{figure*}

\begin{figure*}
\includegraphics[width=18cm,height=15cm]{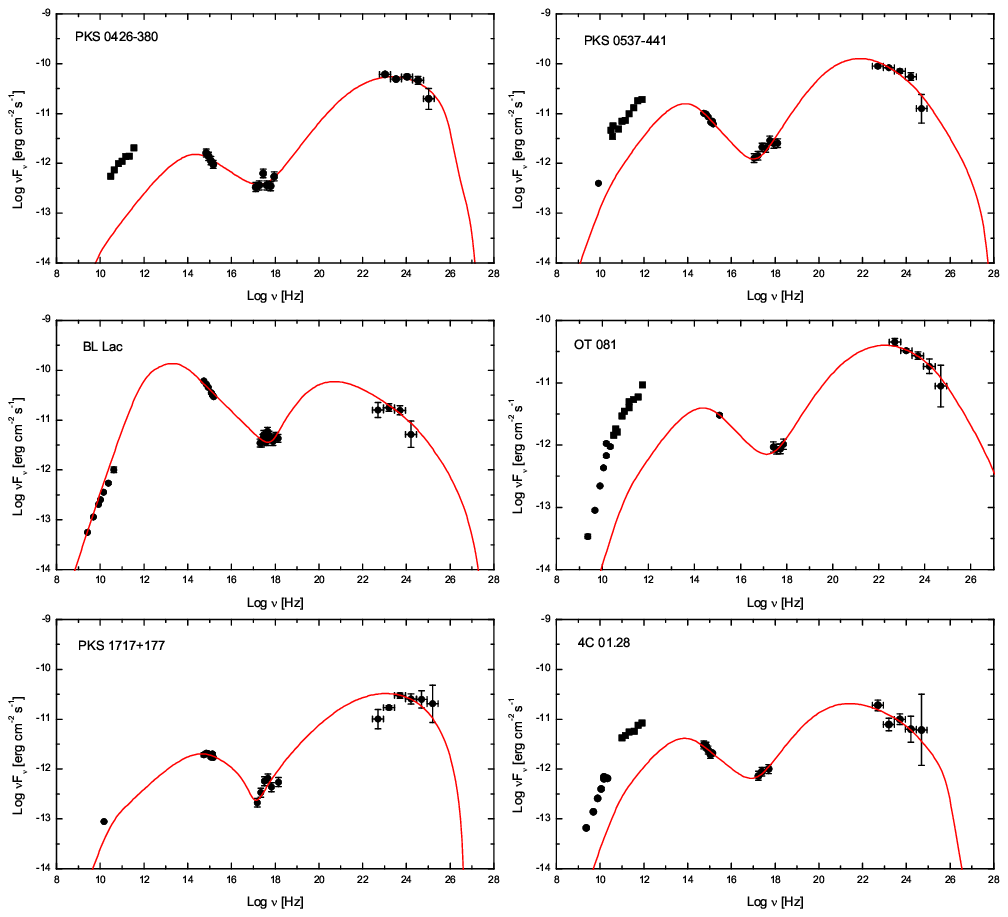}
\caption{Same as in Figure~\ref{f1}, but for LBLs.
}
\label{f3}
\end{figure*}

We compile the broadband SEDs covering from radio, optical, X-ray
to GeV-TeV band from \citet{abdosed,Giommi} and the literature
listed in TeVCat\footnote{http://tevcat.uchicago.edu/} for 22 BL
Lacs having known redshift, including 10 HBLs, 6 IBLs and 6 LBLs
(according to the classification of \citet{Ackermann}). The sources having bad {\it Fermi}
data (only having flux upper limits) in \citet{Giommi} and other literature are also excluded here, although they maybe
have the good optical-UV and X-ray data. The quasi-simultaneous SEDs
including GeV-TeV data of Mrk 421, Mrk 501, PKS 2155-304, PKS
0447-439, 1ES 0414+009, RBS 0413, 1ES 1215+303 and B3 2247+381 are taken from
\citet{abdo421}, \citet{abdo501}, \citet{2155}, \citet{0447}, \citet{0414},
\citet{Aliu12}, \citet{1215} and \citet{2247}, respectively. The
optical-UV data of PKS 0426-380 and 4C 01.28 are taken from
\citet{Giommi}. The rest of SEDs are taken from \citet{abdosed}.
The Plank data taken from \citet{Giommi} are plotted as square.

We apply our model fitting procedure to the SEDs of 22 {\it
Fermi} BL Lacs. The extragalactic background light (EBL) model of \citet{FEBL}
is used to correct the absorption affect. The SEDs of our sources, together with the best-fit one-zone SSC
SEDs, are shown in Figures~\ref{f1}--\ref{f3}. The best-fit model
parameters are listed in Table \ref{para}. It can be found that in general the SEDs of 16 {\it Fermi} HBLs and IBLs
covering from optical to GeV-TeV band are fit well by the
simple one-zone SSC model (see Figures. \ref{f1}--\ref{f2} and Table~\ref{para}) and most of their parameters are
well constrained, except for HBL 1ES 0414+009 and IBL PG 1246+586. It seems that more complex model is needed for
1ES 0414+009 \citep[e.g.,][]{0414}. The first three
months average {\it Fermi} spectrum of IBL PG 1246+586 turns
upward above 5 GeV, and is not a simple power-law, which
evidently can not be explained by the simple one-zone SSC model.
However, its one-year and two-year {\it Fermi} spectrum is
featureless power-law\footnote{http://tool.asdc.asi.it}. Hence, we
do not think it needs more explanations of the turning upward
spectrum in this study. However, the fits to the SEDs of 6 LBLs are bad and
we cannot obtain their meaningful best-fit values (see Figures \ref{f3} and Table~\ref{para}).

\begin{table*}
\caption{The best-fit model parameters and the reduced $\chi^2$.
The first nine sources are HBLs, the second six sources are IBLs,
and the third six sources are LBLs. } \label{para}
\begin{center}
\begin{tabular}{lccccccccccc}
 \hline\hline
Name &
$B$ &
$\delta_{\rm D}$ &
$t_{\rm v,min}$ &
$\gamma^{\prime}_{\rm max}$ &
$\gamma^{\prime}_{\rm b}$ &
$K^{\prime}_{\rm e}$ &
$p_1$ &
$p_2$ &
$\chi^2_{\rm red}$\\
 & (0.01 G) & (10) & ($10^5$ s) & ($10^7$) & ($10^4$) & ($10^{55}$) & & &\\
 \hline
0033-1921 & $4.06\pm1.24$ &  $2.43\pm0.17$ & $2.48\pm1.21$ & $0.07\pm0.01$ & $1.62\pm0.20$ & $0.12\pm0.01$ & $1.83\pm0.08$ & $3.29\pm0.05$ & 1.14\\
0414+009 & $1.30\pm0.58$ &  $2.96\pm1.36$ & $3.54\pm4.31$ & $1.49\pm2.70$ & $12.67\pm1.36$ & $0.04\pm0.02$ & $1.88\pm0.10$ & $3.82\pm0.07$ & 3.96\\
0447-439 & $5.47\pm1.38$ &  $3.63\pm0.08$ & $0.43\pm0.11$ & $0.052\pm0.002$ & $3.18\pm0.29$ & $0.05\pm0.02$ & $2.07\pm0.03$ & $3.96\pm0.17$ & 0.70\\
1013+489 & $5.72\pm0.75$ &  $2.75\pm0.47$ & $0.55\pm0.22$ & $0.08\pm0.04$ & $6.82\pm0.74$ & $0.03\pm0.01$ & $2.03\pm0.04$ & $4.06\pm0.19$ & 2.11\\
2155-304 & $4.89\pm0.66$ &  $1.97\pm0.06$ & $3.47\pm0.52$ & $0.087\pm0.004$ & $3.57\pm0.20$ & $0.011\pm0.002$ & $1.68\pm0.02$ & $3.79\pm0.08$ & 2.48\\
Mrk 421 & $4.23\pm0.41$ & $2.71\pm0.27$ & $0.42\pm0.10$ & $3.73\pm0.81$ & $18.43\pm0.79$ & $0.012\pm0.002$ & $2.13\pm0.02$ & $5.04\pm0.18$ & 1.39\\
Mrk 501 & $2.77\pm0.63$ & $2.99\pm0.70$ & $0.16\pm0.11$ & $0.16\pm0.03$ & $15.81\pm3.10$ & $0.007\pm0.006$ & $2.19\pm0.09$ & $3.12\pm0.04$ & 1.29\\
RBS 0413 & $5.48\pm1.57$ & $2.60\pm0.55$ & $0.23\pm0.11$ & $1.29\pm0.42$ & $9.97\pm1.26$ & $0.0014\pm0.0006$ & $1.93\pm0.07$ & $3.52\pm0.34$ & 1.91\\
1215+303 & $3.49\pm0.17$ & $3.58\pm0.10$ & $0.22\pm0.02$ & $0.27\pm0.01$ & $1.13\pm0.04$ & $0.0031\pm0.0001$ & $1.78\pm0.01$ & $3.61\pm0.04$ & 1.99\\
2247+381 & $5.45\pm1.64$ & $3.62\pm0.05$ & $0.14\pm0.05$ & $0.10\pm0.06$ & $8.87\pm1.96$ & $0.0004\pm0.0002$ & $1.96\pm0.06$ & $4.58\pm0.42$ & 0.54\\
\hline
0048-09 & $6.50\pm5.84$ & $2.50\pm0.28$ & $2.19\pm1.74$ & $0.10\pm0.02$ & $0.52\pm0.04$ & $0.015\pm0.002$ & $1.42\pm0.18$ & $3.72\pm0.08$ & 2.90\\
0716+714 & $5.90\pm1.23$ & $2.71\pm0.47$ & $3.51\pm1.21$ & $0.04\pm0.01$ & $0.92\pm0.10$ & $0.010\pm0.002$ & $1.49\pm0.04$ & $3.88\pm0.07$ & 1.98\\
0851+202 & $4.05\pm2.41$ & $2.40\pm1.10$ & $2.43\pm3.34$ & $0.14\pm0.45$ & $0.26\pm0.10$ & $0.13\pm0.12$ & $1.46\pm0.40$ & $4.65\pm0.16$ & 1.49\\
1058+5628 & $2.20\pm1.14$ & $2.40\pm0.73$ & $1.29\pm0.72$ & $0.06\pm0.03$ & $2.61\pm0.30$ & $0.06\pm0.03$ & $1.93\pm0.05$ & $3.59\pm0.07$ & 1.36\\
1246+586 & $8.82\pm1.89$ & $2.34\pm0.34$ & $3.06\pm0.96$ & $0.40\pm0.02$ & $0.89\pm0.08$ & $0.006\pm0.006$ & $1.43\pm0.03$ & $4.08\pm0.08$ & 1.52\\
W Comae & $4.91\pm0.12$ & $2.70\pm0.13$ & $0.32\pm0.04$ & $0.06\pm0.01$ & $1.94\pm0.09$ & $0.046\pm0.002$ & $2.09\pm0.02$ & $3.65\pm0.04$ & 1.75\\
\hline
0426-380 & $1.08\pm2.42$ & $3.53\pm4.01$ & $0.93\pm1.31$ & $0.47\pm0.02$ & $1.77\pm0.51$ & $0.36\pm0.78$ & $1.78\pm0.51$ & $3.58\pm0.93$ & 2.41\\
0537-441 & $2.12\pm1.55$ & $3.62\pm1.54$ & $1.51\pm1.38$ & $0.38\pm0.40$ & $0.54\pm0.09$ & $0.20\pm0.07$ & $1.56\pm0.13$ & $3.96\pm0.06$ & 5.64\\
1717+177 & $1.79\pm0.20$ & $3.52\pm0.18$ & $0.036\pm0.005$ & $0.013\pm0.003$ & $1.79\pm0.17$ & $0.020\pm0.001$ & $2.12\pm0.04$ & $3.53\pm0.19$ & 3.97\\
BL Lac & $1.86\pm1.89$ & $3.23\pm1.70$ & $0.95\pm0.90$ & $0.11\pm0.09$ & $0.29\pm0.06$ & $0.24\pm0.04$ & $1.84\pm0.18$ & $3.87\pm0.04$ & 4.10\\
OT 081 & $9.82\pm9.80$ & $2.31\pm5.16$ & $0.12\pm0.55$ & $2.00\pm2.10$ & $0.52\pm0.58$ & $0.007\pm0.022$ & $1.75\pm0.66$ & $3.76\pm0.59$ & 1.69\\
4C 01.28 & $10.56\pm19.20$ & $2.47\pm3.42$ & $0.66\pm6.02$ & $0.12\pm0.43$ & $0.30\pm0.18$ & $0.06\pm0.13$ & $1.69\pm0.64$ & $3.70\pm0.32$ & 0.96\\
\hline
  \hline
\end{tabular}
\end{center}
\end{table*}

\section{Results and Discussion}

As mentioned above, this model includes nine parameters,
in which six parameters are used to determine the electron energy
distribution and the others are used to describe the global properties of the
emitting region. Using the above best-fit values of the model
parameters (note that $\gamma_{\rm min}$ is assumed above) listed in
Table~\ref{para}, we can make a statistical analysis on the
electron energy distribution and the properties of the emitting
region. Because the model parameters of LBLs are poorly constrained,
the following statistical analysis are focused on HBLs and IBLs.

\subsection{Relativistic electron distributions}

\begin{figure}
\includegraphics[width=9cm,height=7cm]{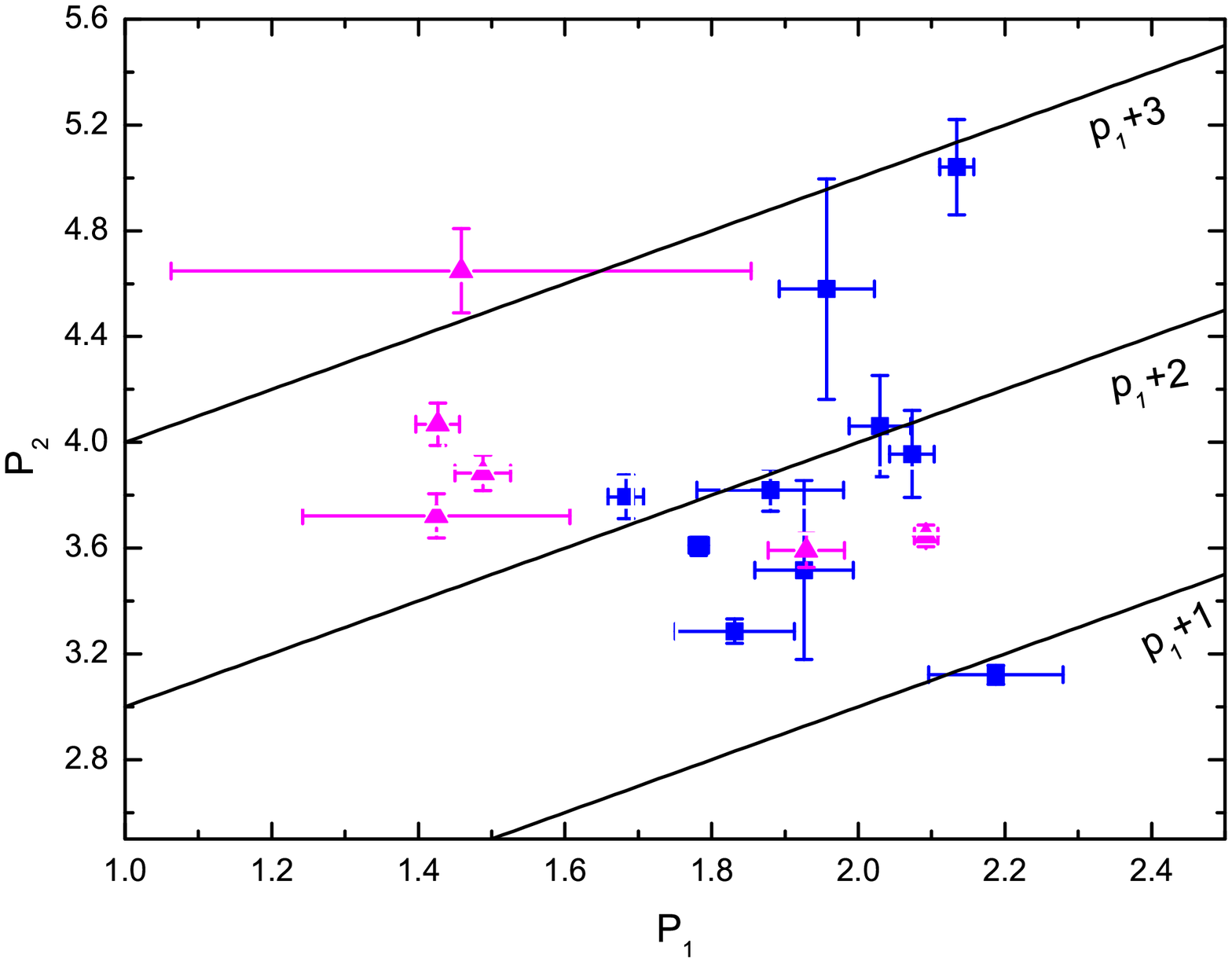}
\caption{ The relativistic electrons spectral indexes $p_1$ vs.
$p_2$. Squares: HBLs, triangles: IBLs.} \label{p1_p2}
\end{figure}

It can be seen that HBLs have distinctly greater
$\gamma^{\prime}_{\rm b}$ than IBLs (see
Table~\ref{para}). For the relativistic electrons spectral indexes
(Figure \ref{p1_p2}), it can be found that the values of $p_2$
cluster around 3.8, while the values of $p_1$ distribute in a
large range (1.4--2.2). Our results show that the values of
($p_2-p_1$) are in the range (1.0--3.0), typically around 2.0. It
is interesting that several IBLs have $p_1<1.6$ and
($p_2-p_1$)$>2.0$, which may imply some clues on the acceleration
process in the jet of a blazar \citep[e.g.,][]{sbacc}.

\subsection{The physical conditions of emitting regions}

\begin{figure}
\includegraphics[width=9cm,height=7cm]{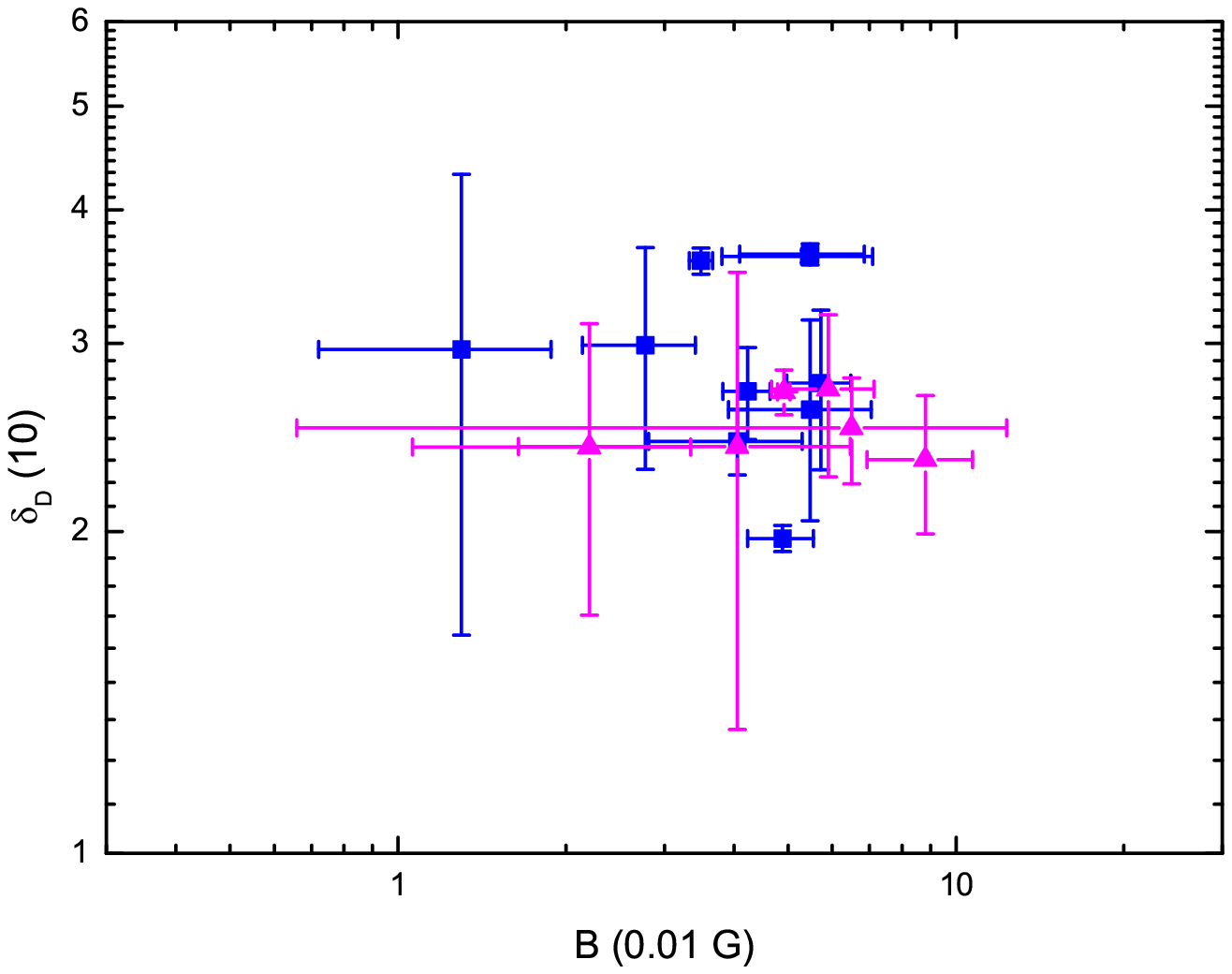}
\caption{The Doppler factor $\delta_{\rm D}$ as a function of $B$. The symbols are same as in Figure~\ref{p1_p2}.
}
\label{B_delta}
\end{figure}

From Table~\ref{para}, it can be found that the derived values of
magnetic field in the emitting regions are in the range
(0.01--0.1) G, which are smaller than the results derived by
\citet{G11,zhang12}, and the values of $\delta_{\rm D}$ are in the
range (20--40), which are consistent with the results based on the
observations \citep{Savolainen}.
According to the SSC model, there
would be relationships between $\delta_{\rm D}$ and $B$ for a given BL
Lac, for example, $B\delta_{\rm D}\propto [\nu^2_{\rm S}/\nu_{\rm
C}](1+z)$ in the Thomson regime and $B/\delta_{\rm D}\propto[\nu_{\rm S}/\nu^2_{\rm
C}]/(1+z)$ in the Klein-Nishina (KN) regime, where $\nu_{\rm S}$ and $\nu_{\rm C}$ are the peak
frequencies of the synchrotron radiation and the inverse Compton
scattering. We have plotted the change of $\delta_{\rm B}$ with
$B$ in Figure~\ref{B_delta} and not found any correlations between $B$
and $\delta_{\rm D}$. The lack of correlations may be due
to the fact that the synchrotron peaks of these objects distribute in a large range.

The obtained minimum variability timescales vary from 1 hour to
$\sim$100 hours (see Table~\ref{para}), and the corresponding
radiuses of emitting blobs are in the range
(0.3--20)$\times10^{16}\ $cm, which are consistent with that of
TeV BL Lacs derived by \citet{zhang12}.

\subsection{Physical conditions of emitting regions vs. relativistic electron distributions}

\begin{figure}
\includegraphics[width=9cm,height=7cm]{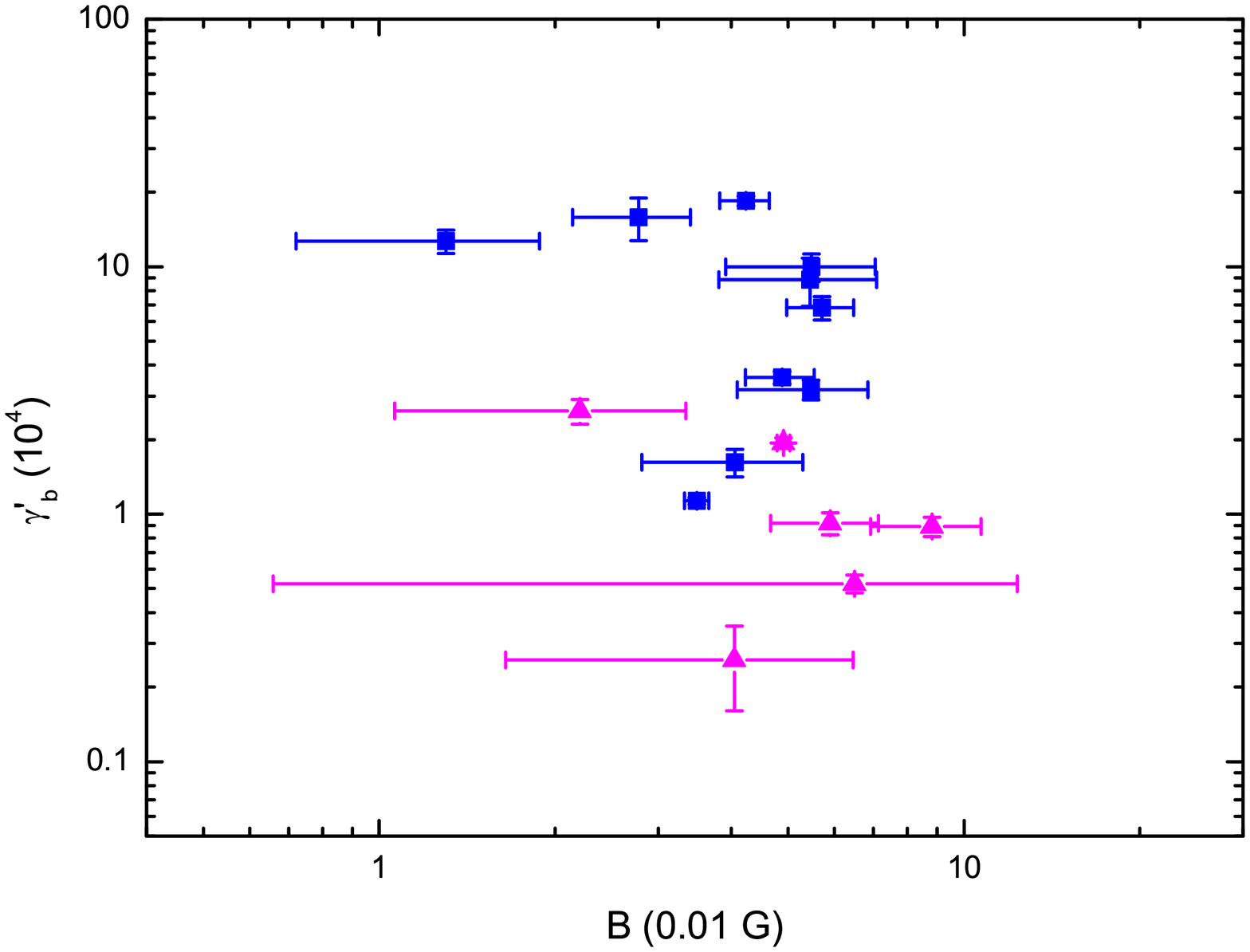}
\caption{$\gamma^{\prime}_{\rm b}$ as a function of $B$.
}
\label{B_gb}
\end{figure}

\begin{figure}
\vskip -0.2cm
\includegraphics[width=9cm,height=7cm]{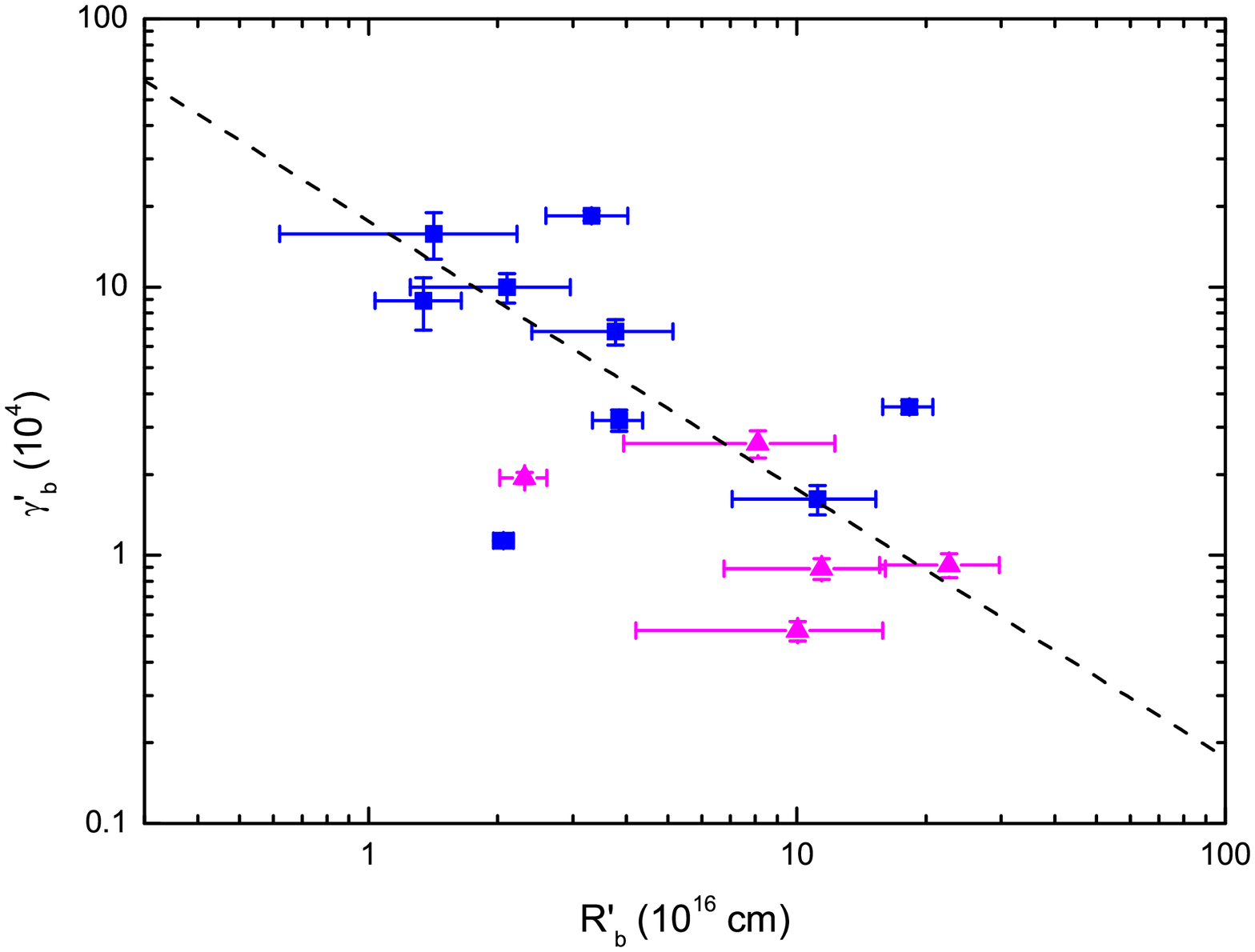}
\vskip -0.6cm
\caption{ $\gamma^{\prime}_{\rm b}$ as a function of $R^{\prime}_{\rm b}$. The dashed line represents the relationship $\gamma^{\prime}_{\rm b}\propto1/[R^{\prime}_{\rm
b}]^{1.0}$ for HBLs and IBLs.
}
\label{gb_Rb}
\end{figure}

\begin{figure}
\vskip -0.2cm
\includegraphics[width=9cm,height=7cm]{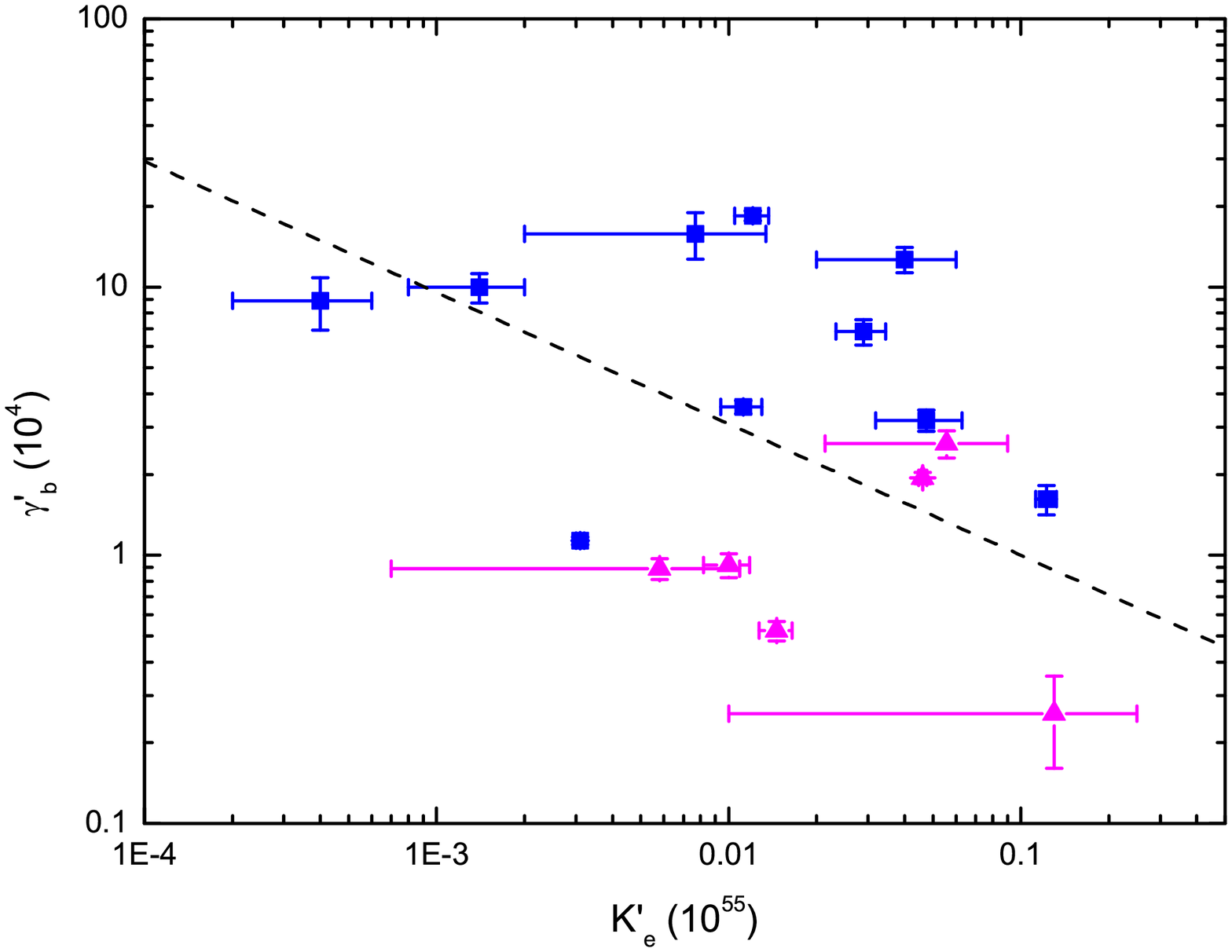}
\vskip -0.6cm
\caption{ $\gamma^{\prime}_{\rm b}$ as a function of $K^{\prime}_{\rm e}$. The dashed line represents the relationship $\gamma^{\prime}_{\rm b}\propto1/[K^{\prime}_{\rm
e}]^{0.49}$ for HBLs and IBLs.
}
\label{gb_ke}
\end{figure}

In the SSC model, $\gamma^{\prime}_{\rm b}\propto
[B\delta]^{-1/2}\nu^{\prime 1/2}_{\rm s}$, which shows that there is a
relationship between $\gamma^{\prime}_{\rm b}$ and $B$, i.e.,
$\gamma^{\prime}_{\rm b}\propto B^{-1/2}$ if $\nu_{\rm s}$ and
$\delta_{\rm D}$ roughly keep constant. On the other hand, the
static electron distribution we used could be considered as a
solution to the electron continuity equation. Such a continuity
equation includes a injection term, a escape term, and a cooling
term (synchrotron + SSC cooling), the continuous injection is
balanced by the cooling and escape \citep{CG99,Li,bott02,Tramacere,yanb}. Hence
$\gamma^{\prime}_{\rm b}$ is obtained when the cooling time
$t_{\rm cool}=3m_{\rm e}c^2/[4\sigma_{\rm
T}c\gamma^{\prime}(U^{\prime}_{\rm B}+U^{\prime}_{\rm syn})]$ is
equal to the escape time $\eta R^{\prime}_{\rm b}/c$, where
$U^{\prime}_{\rm syn}$ is the synchrotron photon field energy
density for IC scattering and $U^{\prime}_{\rm B}=B^2/8\pi$ is the
energy density of magnetic field and $\eta$ is a constant.
Therefore, we have
\begin{equation}
\gamma^{\prime}_{\rm b}\propto1/[(U^{\prime}_{\rm B}+U^{\prime}_{\rm syn})\eta R^{\prime}_{\rm b}]\ .
\label{rb}
\end{equation}
It is clear that $\gamma^{\prime}_{\rm b}$ relies on the cooling process ($U^{\prime}_{\rm syn}$+$U^{\prime}_{\rm B}$) and the escape process $\eta R^{\prime}_{\rm b}$.

In Figures~\ref{f1}-\ref{f2}, it can be roughly estimated that the
ratio of the SSC peak flux to the synchrotron peak flux is less
than unity for HBLs and IBLs, which implies that $U^{\prime}_{\rm
B}/U^{\prime}_{\rm syn}>1$ and the synchrotron cooling is more
important than SSC cooling. Therefore, the relation
$\gamma^{\prime}_{\rm b}\propto B^{-2}$ would be expected for HBLs and
IBLs. However, no correlation between
$B$ and $\gamma^{\prime}_{\rm b}$ is found in our sample (Figure \ref{B_gb}),
which may be caused by the fact that the change of
$\gamma^{\prime}_{\rm b}$ is not caused by the change of $B$.

On the other hand, if $\gamma^{\prime}_{\rm b}$ mainly depends on
the escape, the relation $\gamma^{\prime}_{\rm b}\propto
R^{\prime-1}_{\rm b}$ would be expected (Eq.\ref{rb}).
In Figure \ref{gb_Rb}, we show the distribution of $\gamma^{\prime}_{\rm b}$ vs. $R^{\prime}_{\rm b}$.
The errors of $R^{\prime}_{\rm b}$ are estimated by using a Monte-Carlo method.
The error weights are considered in the correlation analysis.
It is interesting that an anti-correlation is found between
$\gamma^{\prime}_{\rm b}$ and $R^{\prime}_{\rm b}$, i.e.,
$\gamma^{\prime}_{\rm b}\propto1/[R^{\prime}_{\rm
b}]^{1.24\pm0.22}$ for HBLs and IBLs with the correlation coefficient $r=-0.82$ and a chance probability $p=1.21\times10^{-4}$ (Figure \ref{gb_Rb}).
If the change of $\gamma^{\prime}_{\rm b}$ is caused by the escape process,
$\gamma^{\prime}_{\rm b}$ should decrease with increase of $K^{\prime}_{\rm e}$. It's found that $\gamma^{\prime}_{\rm b}$ does inversely correlate with $K^{\prime}_{\rm e}$ with $r=-0.57$ and $p=0.01$ (Figure~\ref{gb_ke}).
The anti-correlations between $\gamma^{\prime}_{\rm b}$ and
$R^{\prime}_{\rm b}$, $K^{\prime}_{\rm e}$ found in our results indicate that for HBLs
and IBLs $\gamma^{\prime}_{\rm b}$ is mainly determined by the escape. Moreover, as discussed above, no inverse and
quadratic correlations between $\gamma^{\prime}_{\rm b}$ and $B$
for HBLs and IBLs also support that the change of $\gamma^{\prime}_{\rm b}$ is
mainly caused by that of the escape instead of $B$.
However, the exact relationship between $\gamma^{\prime}_{\rm b}$ and
$K^{\prime}_{\rm e}$ can not be determined by the current sample.
To achieve it, larger sample with better covered SEDs are required.

\subsection{The physical properties of jets}

\begin{table*}
\caption{The ratios of the energy densities of relativistic
electrons to magnetic fields in the emitting regions and the radiative powers, the jet
powers in the forms of Poynting flux, bulk motion of
electrons and protons (assuming one proton per emitting electron),
as well as the redshifts of the sources. } \centering
\begin{tabular}{lccccccc}
\hline
\hline
Name   &
$U^{\prime}_{\rm e}/U^{\prime}_{\rm B}$ &
$P_{\rm r}$\ (${\rm\ erg\ s^{-1}}$) &
$P_{\rm B}$\ (${\rm\ erg\ s^{-1}}$) &
$P_{\rm e}$\ (${\rm\ erg\ s^{-1}}$) &
$P_{\rm p}$\ (${\rm\ erg\ s^{-1}}$) &
$z$\\
\hline
0033-1912 & 40.15 & $3.19\times10^{44}$ & $4.53\times10^{43}$ & $1.83\times10^{45}$ & $5.32\times10^{45}$ & 0.610\\
0414+009 & 14.79 & $4.24\times10^{43}$ & $2.28\times10^{43}$ & $4.85\times10^{44}$ & $9.62\times10^{44}$ & 0.287\\
0447-439 & 40.81 & $5.29\times10^{43}$ & $2.21\times10^{43}$ & $9.02\times10^{44}$ & $3.52\times10^{45}$ & 0.205\\
1013+489 & 37.60 & $7.25\times10^{43}$ & $1.33\times10^{43}$ & $5.00\times10^{44}$ & $1.61\times10^{45}$ & 0.212\\
2155-304 & 2.41 &  $1.40\times10^{44}$ & $1.47\times10^{44}$ & $2.83\times10^{44}$ & $4.75\times10^{44}$ & 0.116\\
Mrk 421 & 14.54 & $7.42\times10^{42}$ & $5.43\times10^{42}$ & $7.89\times10^{43}$ & $6.65\times10^{43}$ & 0.031\\
Mrk 501 & 269.65 & $1.97\times10^{42}$ & $5.20\times10^{41}$ & $1.41\times10^{44}$ & $5.65\times10^{44}$ & 0.034\\
RBS 0413 & 28.13 & $1.13\times10^{43}$ & $3.36\times10^{42}$ & $9.48\times10^{43}$ & $2.24\times10^{44}$ & 0.190\\
1215+303 & 278.81 & $1.02\times10^{43}$ & $2.5\times10^{42}$ & $6.93\times10^{44}$ & $2.07\times10^{45}$ & 0.130\\
2247+381 & 23.64 & $2.67\times10^{43}$ & $2.63\times10^{42}$ & $6.10\times10^{43}$ & $1.64\times10^{44}$ & 0.130\\
\hline
0048-09 & 32.03 & $9.19\times10^{44}$ & $9.99\times10^{43}$ & $3.20\times10^{45}$ & $7.31\times10^{45}$ & 0.634\\
0716+714 & 1.81 & $3.47\times10^{44}$ & $5.07\times10^{44}$ & $9.19\times10^{44}$ & $1.23\times10^{45}$ & 0.26\\
0851+202 & 135.73 & $5.84\times10^{44}$ & $6.26\times10^{43}$ & $8.50\times10^{45}$ & $1.98\times10^{46}$ & 0.306\\
1058+5628 & 78.23 & $1.65\times10^{43}$ & $6.92\times10^{42}$ & $5.41\times10^{44}$ & $9.84\times10^{44}$ & 0.143\\
1246+586 & 6.40 & $8.54\times10^{44}$ & $2.08\times10^{44}$ & $1.33\times10^{45}$ & $2.36\times10^{45}$ & 0.847\\
W Comae & 150.55 & $1.69\times10^{43}$ & $3.57\times10^{42}$ & $5.37\times10^{44}$ & $1.29\times10^{45}$ & 0.103\\
\hline
0426-380 & $5.42\times10^{4}$ & $2.55\times10^{45}$ & $7.45\times10^{41}$ & $4.04\times10^{46}$ & $1.05\times10^{47}$ & 1.111\\
0537-441 & 2344.32 & $3.20\times10^{45}$ & $1.70\times10^{43}$ & $3.89\times10^{46}$ & $1.08\times10^{47}$ & 0.892\\
1717+177 & $1.63\times10^{5}$ & $1.17\times10^{43}$ & $1.74\times10^{40}$ & $2.83\times10^{45}$ & $1.28\times10^{46}$ & 0.137\\
BL LAC & 237.075 & $1.37\times10^{43}$ & $9.90\times10^{42}$ & $2.35\times10^{45}$ & $3.57\times10^{45}$ & 0.069\\
OT 081 & 2725.29 & $2.20\times10^{44}$ & $7.30\times10^{41}$ & $1.99\times10^{45}$ & $7.26\times10^{45}$ & 0.322\\
4C 01.28 & 369.50 & $1.31\times10^{45}$ & $1.62\times10^{43}$ & $5.98\times10^{45}$ & $2.44\times10^{46}$ & 0.888\\
\hline
\hline
\end{tabular}
\vskip 0.4 true cm
\label{powers}
\end{table*}

\begin{figure}
\includegraphics[width=7cm,height=10cm]{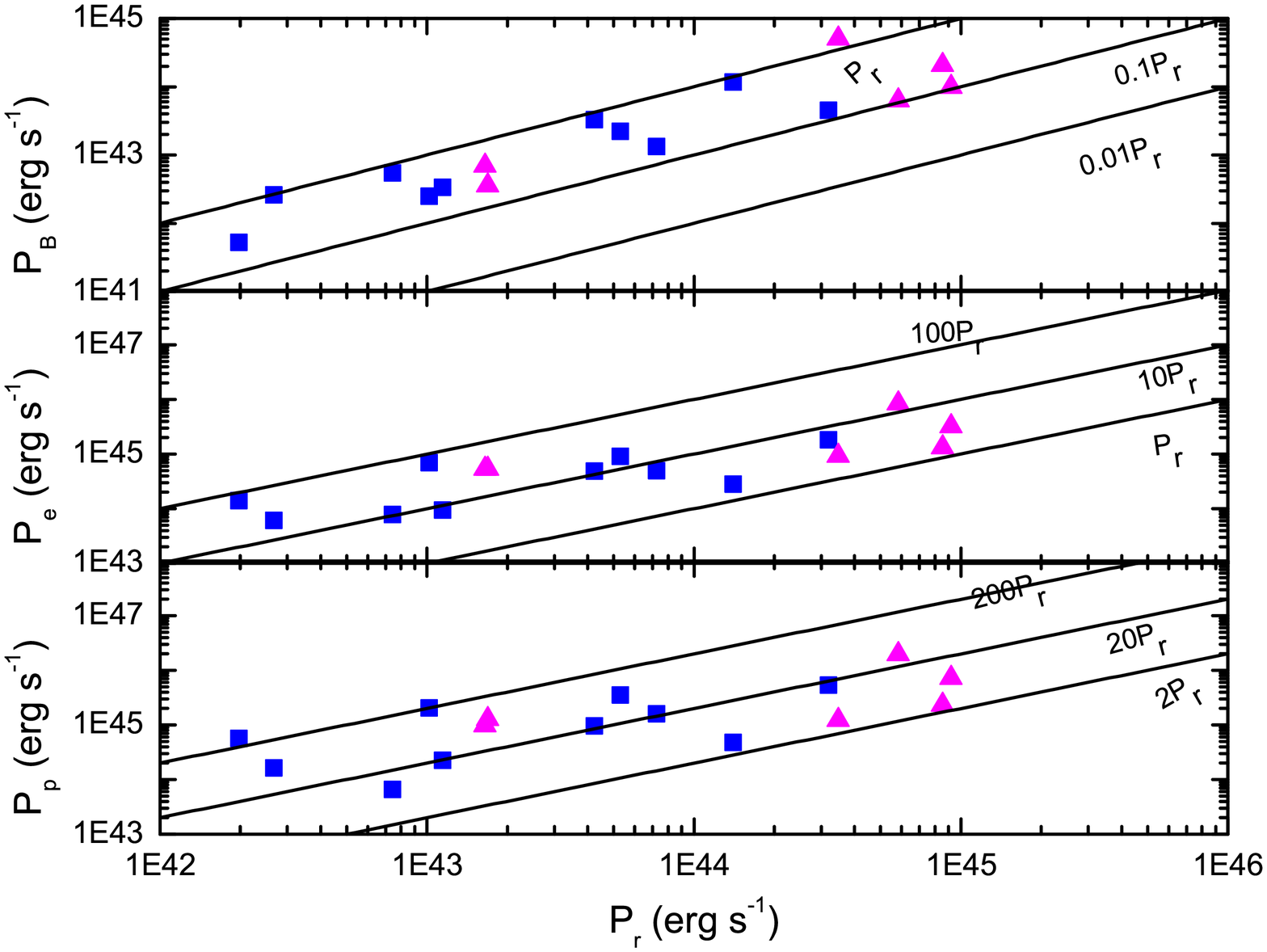}
\caption{Powers in forms of Poynting flux, emitting electrons and
bulk motion of cold protons as functions of the radiative output
$P_{\rm r}$. } \label{pi_pr}
\end{figure}

\begin{figure}
\includegraphics[width=9cm,height=7cm]{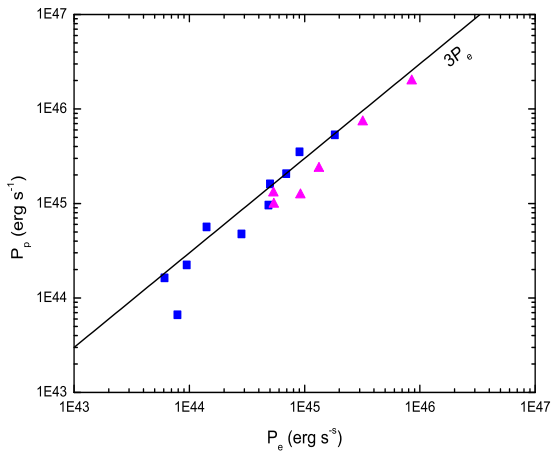}
\caption{ $P_{\rm p}$ as a function of $P_{\rm e}$.
}
\label{pe_pp}
\end{figure}

\begin{figure}
\includegraphics[width=9cm,height=7cm]{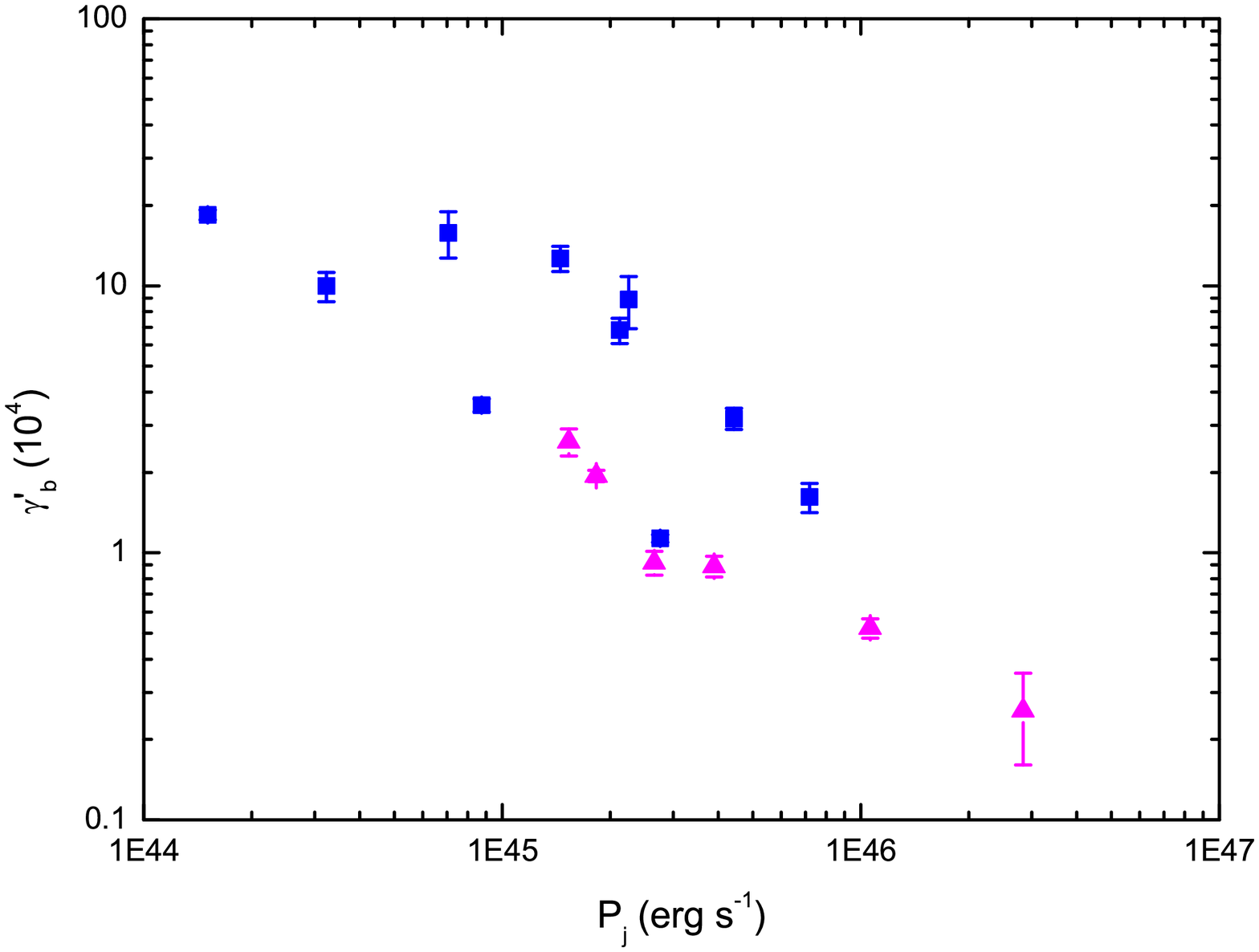}
\caption{ $\gamma^{\prime}_{\rm b}$ versus $P_{\rm j}$.
}
\label{gb_pj}
\end{figure}

After obtaining the values of model parameters, we can estimate
jet power and radiative power. The jet power ($P_{\rm jet}$) is
the sum of Poynting flux power ($P_{\rm B}$) and the powers of
relativistic electrons ($P_{\rm e}$) and protons ($P_{\rm p}$), i.
e., $P_{\rm j}=P_{\rm B}+P_{\rm e}+ P_{\rm p}$ in the stationary
frame of the host galaxy, which are calculated as
\citep{Celotti,Celotti08}
\begin{equation}
P_{\rm i}=\pi R^{\prime 2}_{\rm b}\Gamma^2U^{\prime}_{\rm i}c\ ,
\end{equation}
where $U^{\prime}_{\rm i}$ (i= e, B, p) are the energy densities
associated with  the emitting electrons $U^{\prime}_{\rm e}$,
magnetic field $U^{\prime}_{\rm B}$, and protons $U^{\prime}_{\rm
p}$ in the comoving frame, respectively. We calculate
$U^{\prime}_{\rm p}$ by assuming one proton per emitting electron,
then $U^{\prime}_{\rm p}=U^{\prime}_{\rm e}(m_{\rm p}/m_{\rm
e})/\langle\gamma^{\prime}\rangle$ \citep{Celotti08}, where
$\langle\gamma^{\prime}\rangle=\frac{\int N^{\prime}_{\rm e}(\gamma^{\prime})\gamma^{\prime}d\gamma^{\prime}}{\int N^{\prime}_{\rm e}(\gamma^{\prime})d\gamma^{\prime}}$ is the average energy of
relativistic electrons. Here, we take the bulk Lorentz factor
$\Gamma=\delta_{\rm D}$. On the other hand, for the radiative
power, the energy density can be expressed as $U^{\prime}_{\rm
r}=L^{\prime}/(\pi R^{\prime 2}_{\rm b}c)$, so it reads
\citep{Celotti08}
\begin{equation}
P_{\rm r}=L^{\prime}\Gamma^2\approx L\frac{\Gamma^2}{\delta^4_{\rm
D}}\ ,
\end{equation}
where $L$ is total non-thermal luminosity.

It should be pointed out that the estimate of the proton kinetic
power and relativistic electron's power are dependent on the value
of $\gamma^{\prime}_{\rm min}$, which however is historically
poorly constrained by the modelling, especially for low-power BL
Lacs \citep[e.g.,][]{Celotti08}. Due to the synchrotron
self-absorption, the
radio emission cannot be used to constrain the value of
$\gamma^{\prime}_{\rm min}$. However, the {\it Swift}-BAT
observed hard X-ray data and {\it Fermi} GeV data at low energies
could place constraints on $\gamma^{\prime}_{\rm min}$ in some
degree. From our SED modeling results (Figures~\ref{f1}--\ref{f3}),
it can be seen that the hard X-ray data and {\it Fermi} GeV data below 1
GeV can be fitted well. Therefore, we would like to believe that the
jet kinetic powers we derived here are creditable.

Table~\ref{powers} lists the ratios of the energy densities of
relativistic electrons to magnetic fields in the emitting regions
$U^{\prime}_{\rm e}/U^{\prime}_{\rm B}$, and the powers carried by
the jets in the forms of radiations, Poynting flux, relativistic
electrons and protons (assuming one proton per emitting electron).

From the results given by Table~\ref{powers}, we can see $P_{\rm
e}>P_{\rm B}$ and $U^{\prime}_{\rm e}>U^{\prime}_{\rm B}$ in our sample,
thus these jets are particle-dominated. In Figure \ref{pi_pr}, we
show the changes of $P_{\rm B}$, $P_{\rm e}$, and $P_{\rm p}$ with
radiative power $P_{\rm r}$. It can be seen that $P_{\rm r}>P_{\rm
B}$ and $P_{\rm r}/P_{\rm e}\sim0.01-0.8$. The former means that
the Poynting flux cannot account for the observed radiation and
the latter indicates that a large fraction of the relativistic electron
power would be used to produce the observed radiation. Comparing middle
panel with bottom panel in Figure \ref{pi_pr}, we can see $P_{\rm
p}>P_{\rm e}$. In fact, we find $P_{\rm p}\approx3 P_{\rm e}$ in
Figure \ref{pe_pp}. As mentioned above, it implies that only a small fraction of the jet power is dissipated into
the observed radiation, $P_{\rm r}/P_{\rm j}\sim$ 1--30 percent.

Furthermore, the relation $P_{\rm p}\approx3 P_{\rm e}$ in Figure \ref{pe_pp}
indicates that the average energy of relativistic electrons $\langle\gamma^{\prime}\rangle\approx610$,
about one third of $m_{\rm p}/m_{\rm e}$, which may indicate that only a very small fraction $\eta_{\rm e}\sim1\%$ of the energy dissipated
in the shock is picked up by the electrons, where $\eta_{\rm e}=\frac{\langle\gamma^{\prime}\rangle}{\Gamma}(m_{\rm e}/m_{\rm p})$ \citep{Giannios} and $\Gamma=25$ is used.

Compared to the result $P_{\rm r}\sim P_{\rm e}$ derived by \citet{Celotti08},
the relation $P_{\rm r}<P_{\rm e}$ derived here is due to the fact
that our sample are the low power BL lacs. Due to the
efficient cooling of electrons and the result we derived that $P_{\rm r}/P_{\rm e}$ can be $\sim 0.8$,
it may be safe to suggest
that an additional energy reservoir of cold hadrons is needed to
accelerate electrons \citep[e.g.,][]{Celotti08}.

\subsection{The blazar sequence and implications on the differences between HBLs/IBLs and LBLs}

Our results show that there is an anti-correlation between
$\gamma^{\prime}_{\rm b}$ and $P_{\rm j}$ for HBLs and IBLs, i.e.,
$\gamma^{\prime}_{\rm b}\propto P^{-(0.92\pm1.78)}_{\rm j}$ with $r=-0.89$ and $p=6.10\times10^{-6}$
(Figure \ref{gb_pj}). As mentioned in \citet{Celotti08}, this result
is consistent with the prediction of the blazar sequence
\citep{f98,G98,G08} and is usually explained as that the radiative
cooling is stronger in more powerful blazar. As we
discussed above, it is the escape
of relativistic electron which strongly affect the variation of
$\gamma^{\prime}_{\rm b}$ for HBLs and IBLs modeled here.

As mentioned in Section 3, the one-zone SSC model fails to fit the SEDs of LBLs.
Furthermore, from Table \ref{powers} it can be found that LBLs have extreme values of $U^{\prime}_{\rm
e}/U^{\prime}_{\rm B}$ (400--$10^5$) in the frame of a one-zone SSC model.
On the other hand, the simple
one-zone SSC model predicts that the synchrotron emission peak
frequencies of three LBLs (PKS 0426-380, PKS 1717+177, OT 081) are
larger than $10^{14}\ $Hz, which indicate that we overestimated the synchrotron peak
frequencies in the SSC model. It is therefore suggested again
that the one-zone SSC model is not the right model accounting for
the multi-wavelength radiations from LBLs. With a large {\it Fermi}
blazar sample, \citet{Fan} suggested that the spectral index
properties of LBLs are similar with that of FSRQs.
Comparing Figure \ref{f3} with Figures \ref{f1} and \ref{f2}, it can be found
that the ratios of the Compton to the synchrotron peak energy
fluxes of LBLs are greater than those of HBLs and IBLs, and then
LBLs are Compton dominated. As suggested in \citet{finke13}, the
Compton dominance is an more intrinsic indicator for blazar
sequence.

\section{Conclusions}

We have modeled the quasi-simultaneous SEDs of 22 {\it Fermi} BL
Lacs with a one-zone SSC model. We use a $\chi^2$-minimization
procedure to obtain the best-fit model parameters and their errors,
then the jet powers and the radiative powers are calculated.

Based on the derived results, we firstly discussed their
implications on the physical processes in the emitting blobs. It
can be found that there is no correlation between
$\gamma^{\prime}_{\rm b}$ and $B$. This lack of correlation was
also found by \citet{zhang12} in a TeV BL Lacs sample and by
\citet{man} in different states of Mrk 501. It seems that this
lack of correlation is common in BL Lacs especially in HBLs and IBLs.
Moreover, It can be found from our results that $\gamma^{\prime}_{\rm b}$ is inversely correlated
with $R^{\prime}_{\rm b}$ as well as $K^{\prime}_{\rm e}$ for HBLs and
IBLs. These results indicate that in the emitting blobs of
HBLs/IBLs where the synchrotron cooling is more important than IC cooing,
$\gamma^{\prime}_{\rm b}$ is mainly determined by the escape process.

Secondly, we concerned the powers of the {\it Fermi} BL Lacs jets.
Our results confirm that the jet is energetically dominated by the
proton component and only a small fraction of the jet power is
transformed into radiation if there is one proton for a emitting
electron \citep[e.g.,][]{Celotti08}. Moreover, based on the
physical properties of relativistic jets, our results
confirm that HBLs/IBLs are different from LBLs again
\citep[e.g.,][]{Fan}.

\section*{Acknowledgments}
We thank the anonymous referee for helpful comments.
We thank P. Giommi, D. Paneque, D. Sent\"{u}rk, A. Smith and E. Lindfors for sending us some data sets we used here.
We acknowledge the support of Yunnan University's
Science Foundation for graduate students under grant No. YNUY201260 and the Science Foundation for graduate students of Provincial Education Department of Yunnan under grant No. 2013J071.
This work is partially supported by the Science Foundation of Yunnan Province under a grant 2009 OC.

\bibliography{refernces}

\begin{thebibliography}{}

\bibitem[Abdo et al.(2010)]{abdosed}Abdo A., A., et al., 2010, ApJ, 716, 30

\bibitem[Abdo et al.(2011a)]{abdo3c66a}Abdo A. A., Ackermann M., Ajello M., et al. 2011a, ApJ, 726, 43

\bibitem[Abdo et al.(2011b)]{abdoBL}Abdo A. A., Ackermann M., Ajello M., et al. 2011b, ApJ, 730, 101

\bibitem[Abdo et al.(2011c)]{abdo421}Abdo, A. A., Ackermann, M., Ajello, M., et al. 2011c, ApJ, 736, 131

\bibitem[Abdo et al.(2011d)]{abdo501}Abdo, A. A., Ackermann, M., Ajello, M., et al. 2011d, ApJ, 727, 129

\bibitem[Ackermann et al.(2011)]{Ackermann}Ackermann M., et al., 2011, ApJ, 743, 171

\bibitem[Aharonian et al.(2009)]{2155}Aharonian F., et al. (H.E.S.S. Collaboration) 2009, ApJ, 696, L150

\bibitem[Aleksi\'{c}  et al.(2012a)]{1215}Aleksi\'{c}, J., et al. 2012a, A\&A, 544, 142

\bibitem[Aleksi\'{c}  et al.(2012b)]{2247}Aleksi\'{c}, J., et al. 2012b, A\&A, 539, 118


\bibitem[Aliu et al.(2012a)]{0414}Aliu E., et al. 2012a, ApJ, 755, 118

\bibitem[Aliu et al.(2012b)]{Aliu12}Aliu E., et al. 2012b, ApJ, 750, 94

\bibitem[Andrae et al.(2010)]{caveat}Andrae R., Schulze-Hartung T., \& Melchior P. 2010, arXiv:1012.3754

\bibitem[B\"{o}ttcher \& Chiang(2002)]{bott02}B\"{o}ttcher M. \& Chiang J.2002, ApJ, 581, 127

\bibitem[B\"{o}ttcher(2007)]{bott07} B\"{o}ttcher  M. 2007, Ap\&SS, 309, 95

\bibitem[Celotti \& Fabian(1993)]{Celotti}Celotti A., \& Fabian A. C. 1993, MNRAS, 264, 228

\bibitem[Celotti \& Ghisellini(2008)]{Celotti08}Celotti A., \& Ghisellini G. 2008, MNRAS, 385, 283

\bibitem[Chiaberge \& Ghisellini(1999)]{CG99}Chiaberge M. \& Ghisellini G. 1999, MNRAS, 306, 551

\bibitem[Dermer \& Schlickeiser(1993)]{dermer93}Dermer C. D., \& Schlickeiser R. 1993, ApJ, 416, 458

\bibitem[Dermer et al.(2009)]{dermer09}Dermer C. D., Finke,  J. D., Krug  H., \& B\"{o}ttcher  M. 2009, ApJ, 692, 32

\bibitem[Dermer et al.(2012)]{dermer12}Dermer C. D., Murase K., Takami H. 2012, ApJ, 755, 147

\bibitem[Dimitrakoudis et al.(2012)]{hadron12}Dimitrakoudis S., Mastichiadis A., Protheroe R. J., Reimer A. 2012, A\&A, 546, A120

\bibitem[Fan et al.(2012)]{Fan}Fan J. H., Yang J. H., Yuan Y. H., Wang J., Gao Y. 2012, ApJ, 761, 125

\bibitem[Finke et al.(2008)]{finke08}Finke J. D., Dermer C. D., \& B\"{o}ttcher, M. 2008, ApJ, 686, 181

\bibitem[Finke(2013)]{finke13}Finke J. D. 2013, ApJ, 763, 134

\bibitem[Fossati et al.(1998)]{f98}Fossati G., Maraschi L., Celotti A., Comastri A., Ghisellini G. 1998, MNRAS, 299, 433

\bibitem[Franceschini et al.(2008)]{FEBL}Franceschini A., Rodighiero G., \& Vaccari M. 2008, A\&A, 487, 837

\bibitem[Giannios \& Spitkovsky(2009)]{Giannios}Giannios D. \& Spitkovsky A. 2009, MNRAS, 400, 330

\bibitem[Ghisellini et al.(1998)]{G98}Ghisellini G., Celotti A., Fossati G., Maraschi L., Comastri, A. 1998, MNRAS, 301, 451

\bibitem[Ghisellini \& Tavecchio(2008)]{G08}Ghisellini G. \& Tavecchio F. 2008, MNRAS, 387, 1669

\bibitem[Ghisellini \& Tavecchio(2009)]{G09}Ghisellini G. \& Tavecchio F. 2009, MNRAS, 397, 985

\bibitem[Ghisellini et al.(2009)]{G09jet}Ghisellini G., Tavecchio F., Ghirlanda G. 2009, MNRAS, 399, 2041

\bibitem[Ghisellini et al.(2010)]{G10}Ghisellini G., Tavecchio F., Foschini L., Ghirlanda G., Maraschi L., Celotti, A. 2010, MNRAS, 402, 497

\bibitem[Ghisellini et al.(2011)]{G11}Ghisellini G., Tavecchio F., Foschini L., Ghirlanda G. 2011, MNRAS, 414, 2674

\bibitem[Giommi et al.(2012)]{Giommi}Giommi, P. et al. 2012, A\&A, 541, 160

\bibitem[Jones(1968)]{Jones68}Jones, F. C. 1968, Phys. Rev., 167, 1159

\bibitem[Kirk et al.(1998)]{Kirk}Kirk J. G., Rieger, F. M., Mastichiadis, A. 1998, A\&A, 333, 452


\bibitem[Li \& Kusunose(2000)]{Li}Li H., \& Kusunose M. 2000, ApJ, 536, 729

\bibitem[Mannheim(1993)]{Mannheim}Mannheim, K., 1993, A\&A, 269, 67

\bibitem[Maraschi et al.(1992)]{Maraschi}Maraschi, L., Ghisellini, G., \& Celotti, A. 1992, ApJL, 397, L5

\bibitem[Mankuzhiyil et al.(2011)]{man11}Mankuzhiyil N., Ansoldi S., Persic M., \& Tavecchio F. 2011, ApJ, 733, 14

\bibitem[Mankuzhiyil et al.(2012)]{man}Mankuzhiyil N., Ansoldi S., Persic M. et al. 2012, ApJ, 753, 154

\bibitem[M\"{u}cke et al.(2003)]{Mucke}M\"{u}cke, A., Protheroe, R. J., Engel, R. et al. 2003, APh, 18, 593

\bibitem[Prandini et al.(2012)]{0447}Prandini E.,  Bonnoli G.,  Tavecchio F. 2012, A\&A, 543, 111

\bibitem[Press et al.(1992)]{press}Press W.H., et al. 1992, Numerical Recipes (Cambridge: Cambridge University Press)

\bibitem[Rees(1967)]{rees}Rees M. J. 1967, MNRAS, 137, 429


\bibitem[Savolainen et al.(2010)]{Savolainen}Savolainen T., Homan D. C., Hovatta T. et al., 2010, A\&A, 512, 24

\bibitem[Sikora et al.(1994)]{Sikora94}Sikora M., Begelman M. C., \& Rees M. J. 1994, ApJ, 421, 153


\bibitem[Summerlin \& Baring(2012)]{sbacc}Summerlin E. J., \&  Baring M. G. 2012, ApJ, 745, 63

\bibitem[Tavecchioet et al.(1998)]{tav98}Tavecchio F., Maraschi L., Ghisellini G. 1998, ApJ, 509, 608


\bibitem[Tramacere et al.(2011)]{Tramacere}Tramacere A., Massaro E., Taylor A. M. 2011, ApJ, 739, 66

\bibitem[Yan et al.(2012a)]{yan}Yan D. H., Zeng H. D., Zhang L. 2012a, PASJ, 64, 80

\bibitem[Yan et al.(2012b)]{yanb}Yan D. H., Zeng H. D., Zhang L. 2012b, MNRAS, 424, 2173

\bibitem[Yan et al.(2013)]{yan2}Yan D. H., Zhang L., Yuan Q., Fan Z. H., Zeng H. D. 2013, ApJ, 765, 122

\bibitem[Zhang et al.(2012)]{zhang12}Zhang J., Liang E. W., Zhang S. N., Bai J. M. 2012, ApJ, 752, 157

\end{thebibliography}

\end{document}